\documentclass[12pt]{article} 
\pdfoutput=1
\usepackage{amssymb,graphicx}
\usepackage{epstopdf}
\usepackage{amsmath,amsfonts}
\usepackage{epsfig} 
\usepackage{graphicx,graphics}

\usepackage{xcolor}

\usepackage{authblk}

\begin{document}

\date{}
\sloppy
\title{\bf\vskip-1cm Dark Matter and Baryon Asymmetry from the very Dawn of Universe}
\author[1,2]{Eugeny Babichev}
\author[3,4]{Dmitry Gorbunov}
\author[5]{Sabir Ramazanov}
\affil[1]{Laboratoire de Physique Th\'eorique, CNRS, \protect\\ Univ. Paris-Sud, Universit\'e Paris-Saclay, 91405 Orsay, France}
\affil[2]{UPMC-CNRS, UMR7095, Institut d'Astrophysique de Paris, \protect\\ ${\mathcal{G}}{\mathbb{R}}\varepsilon{\mathbb{C}}{\mathcal{O}}$, 
98bis boulevard Arago, F-75014 Paris, France}
\affil[3]{Institute for Nuclear Research of the Russian Academy of Sciences, \protect\\ 60th October Anniversary prospect 7a, Moscow 117312, Russia}
\affil[4]{Moscow Institute of Physics and Technology, \protect\\ Institutsky per. 9, Dolgoprudny 141700, Russia}
\affil[5]{CEICO-Central European Institute for Cosmology and Fundamental Physics, \protect\\ Institute of Physics of the Czech Academy of Sciences,
\protect\\ Na Slovance 2, 182 21 Prague 8, Czech Republic}
 
 \renewcommand\Affilfont{\itshape\small}

{\let\newpage\relax\maketitle}

\begin{abstract}

We propose a universal mechanism of producing dark matter and
baryon (lepton) charge at the stage of the quasi-de Sitter expansion
of the Universe---inflation. 
The key ingredient of the mechanism is a linear coupling of the field, responsible for generation of dark matter or baryon (lepton) charge, 
to a function of the inflaton. 
During inflation this induces almost constant force dragging the corresponding field to the non-zero value. This force explicitly breaks quantum numbers associated with dark matter/baryon abundance at later stages. 
As a particular realization of the mechanism we introduce a super-heavy complex scalar field with the mass larger than the Hubble rate during the last e-folds of inflation. The global $U(1)$-symmetry is violated due to the linear coupling of the phase 
of the complex scalar to the inflaton. 
The symmetry breaking leads to the generation of a non-zero Noether charge. 
The latter is directly related to the dark matter abundance, or, alternatively, can be converted into baryon asymmetry, if the complex scalar carries the baryon charge.

\end{abstract}

\section{Introduction}

The earliest yet hypothetical stage of the Universe
evolution---inflation (almost exponential expansion)---has been
suggested to solve simultaneously the so called initial condition
problems of the Hot Big Bang theory: singularity, flatness,
horizon problems and the Universe overclosure by heavy
relics~\cite{Starobinsky:1980te, Guth:1980zm, Linde:1981mu,
  Albrecht:1982wi, Linde:1983gd}. By itself, this idea is
unfalsifiable, unless we observe the spatial curvature, which is
predicted to be exponentially small. However, if inflation is driven by a scalar field called inflaton, quantum fluctuations of
the latter evolve into classical field perturbations, which serve as primordial matter perturbations sourcing the
large-scale structure formation and anisotropy of the cosmic microwave
background in the late Universe~\cite{Mukhanov:1981xt}. In this manner, inflation can solve another problem of the Hot Big Bang theory and becomes testable through observed parameters of the matter power spectrum, which are determined by the inflaton dynamics.

In this paper we put forward an idea that two other problems of the
Hot Big Bang theory --- dark matter (DM) and baryon asymmetry (BA) of the
Universe --- can also be solved at the inflationary stage by the
same source, inflaton. From the analysis of cosmological data we know
that the baryon component must be asymmetric at least from
the epoch of the primordial nucleosynthesis (temperatures above
1\,MeV), while the DM component must be present in the
Universe well before the plasma temperature drops to eV-scale. Usually,
two different mechanisms are invoked to produce DM and
generate the asymmetry~\cite{Dolgov:1991fr, Rubakov:1996vz, Trodden:1998ym, Gorbunov:2011zz} at different stages starting from the Universe
reheating. We suggest that both processes can happen even earlier, at
inflation, and moreover, they can be triggered by one and the same
source (inflaton in our case), which is arguably natural given the
order-of-magnitude coincidence between DM and baryon
contribution to the energy density of the present Universe.

The very idea seems to be in gross contradiction with our expectations about the properties of the Universe 
after inflation. Indeed, the latter gets
rid of all the rubbish from previous stages and makes a huge part of the 
Universe, formed by many Hubble-size patches, flat, causal and
empty. However, the inflaton perturbations provide with one example of
the remnants from inflationary epoch which do not only survive till today,
but are extremely important to shape realistically the late
Universe. Encouraged by this example we introduce the interactions
between the inflaton and DM, as well as the inflaton and a field carrying baryon charge. The interactions are designed in such a way that the inflaton plays a
role of external source of DM and baryons. Being almost
constant during inflation, the source provides a non-trivial attractor for DM field and the field carrying  baryon charge. 
The latter property is
welcome and inherent in the inflaton nature, which washes out any
traces of the initial conditions at the beginning of inflation.
The attractor solutions are governed by the chosen coupling constants
and can explain both DM and BA at
this earliest stage of the Universe evolution.

As for the concrete realization of the idea, we consider a massive
complex scalar field charged under a global $U (1)$-symmetry
group. We will be mainly (but not only) interested in the super-heavy
field with the mass $M$ larger than the Hubble rate $H$ during
inflation, $M \gtrsim H$.  Were the $U(1)$-symmetry exact at all the
times, any pre-existing $U(1)$-charge would decay exponentially fast
in the course of the inflationary expansion, and no charge would be left at the hot epoch (Section~2). 
The problem can be
circumvented by breaking the $U(1)$-symmetry (Section~3). 
To achieve this, we couple the phase $\varphi$ of the complex scalar field directly to the inflaton, i.e.,
$$\sim \varphi \cdot  T_{infl},$$
where $T_{infl}$ is the stress-energy tensor of the inflaton. This coupling generates the non-zero Noether charge, and keeps it up while inflation proceeds.

We study cosmological perturbations in this model in Section~4. The adiabatic mode of the
complex scalar perturbations is nearly constant beyond the horizon
during inflation. In the case of super-heavy field, $M \gtrsim H$ (which we focus on in the bulk of the paper), 
the isocurvature mode is decaying exponentially fast. On the contrary, for 
masses $M \ll H$  (considered in the Appendix), large isocurvature perturbations are generated, unless the scale of
inflation is very low or the complex field is ultra-light. One can
avoid these constraints by the price of introducing the non-minimal
coupling to gravity. The latter translates into the large effective
mass of the complex field during inflation and allows to suppress the isocurvature mode. We consider this option in Section~5.

So far, the nature of the complex scalar has been irrelevant for our discussion. Its physical consequences may be twofold after inflation. 
On one hand, one may assume that the non-zero baryon charge is attributed to the complex scalar. Then its Noether
charge can be transmitted to the observed BA {\it \`a la}
the Affleck--Dine mechanism~\cite{Affleck:1984fy}. On the other hand, 
the complex scalar can be viewed as DM. In the case of super-heavy masses, $M
\gtrsim H$, the energy density of DM is directly related to the Noether
charge density, which redshifts as $1/a^3$. 
Each of the components of the complex field experiences coherent
oscillations around the minimum, and DM behaves cosmologically as the pressureless
perfect fluid.

In the rest of the paper we refer to the complex field as DM, unless the opposite is stated, although in most cases the discussion remains essentially the same for both DM and the field carrying baryon charge. 
Indeed, we are primarily interested in the dynamics of the complex scalar during inflation and shortly afterwards. Then, for the inflaton it is fairly irrelevant, 
whether the complex scalar is attributed to DM or it is used to generate BA. There is a quantitative difference, however, because 
producing the right amount of DM and BA requires different coupling constants to the inflaton. We will get back to this point in Section 3.3, when making the relevant estimates.

The paper is organized as follows. In Section~2 as a warm up we briefly consider the case of a non-interacting complex scalar field at inflation (with more details given in the Appendix).  
In Section~3, we discuss the production of the complex scalar through the $U(1)$-symmetry breaking coupling of its phase to the inflaton. We study linear adiabatic and isocurvature perturbations of the complex scalar in Section~4. 
In Section~5, we suggest a way to avoid large isocurvature
perturbations in the case of light field. We discuss obtained results in Section~6.

\section{Warm up: Non-interacting complex scalar during inflation}

We start with the action of the canonical complex scalar field, 
\begin{equation}
\label{lambdadyn}
S_\Psi=\int d^4 x \sqrt{-g} \left[\frac{1}{2} |\partial_{\mu} \Psi |^2-V(|\Psi|)\right]
\; ,
\end{equation}
where $V(|\Psi|)$ is potential. In the bulk of the paper we assume that the scalar field is minimally coupled to gravity\footnote{The case of the non-minimal coupling is considered in Section~5.}, 
\begin{equation}
\label{actionEH}
	S_{EH}= \frac{M_{Pl}^2}{16\pi}\int d^4 x \sqrt{-g} R\,,
	\nonumber
\end{equation}
where $M_{Pl}$ is the Planck mass related to the Newton constant $G= M_{Pl}^{-2}$. 
We further neglect self-interaction of the scalar field. This is necessary once we treat the scalar field as DM. 
On the other hand, when $\Psi$ is viewed as a field generating BA, such an assumption is allowed although not necessary. 
Nevertheless, in order to keep the discussion homogeneous, we take the potential $V$ to be quadratic for both DM and the field generating BA, 
\begin{equation}
\nonumber 
V(|\Psi|)=\frac{M^2|\Psi|^2}{2} \; .
\end{equation}
In terms of amplitude $\lambda$ and phase $\varphi$ of the complex scalar $\Psi=\lambda e^{i\varphi}$ , action~\eqref{lambdadyn} takes the form, 
\begin{equation}
\label{actionlambdavarphi}
S_\Psi=\int d^4 x \sqrt{-g} \left[ \frac{1}{2} (\partial_{\mu} \lambda )^2+\frac{1}{2} \lambda^2 (\partial_{\mu} \varphi )^2 -V(\lambda) \right] 
\; .
\end{equation} 
We consider the situation, where the field $\Psi$ is a spectator at
very early times, and the cosmological evolution is dominated by the
slowly rolling canonical inflaton field $\phi$. Thus, the Friedmann equation reads, 
\begin{equation}
\nonumber 
H^2\approx\frac{8\pi}{3 M^2_{Pl}} U \; ,
\end{equation}
where $U$ is the inflaton potential. The derivative of the Hubble rate is given by 
\begin{equation}
\nonumber 
\dot{H}=-4\pi G \dot{\phi}^2 \; .
\end{equation}

Our main goal in the present work is to study the production of DM (or BA) through the interaction with the inflaton introduced in the next Section. 
Beforehand, it is instructive to review the case, when no interaction is present. 
The model~(\ref{lambdadyn}) possesses global $U(1)$-symmetry. The associated Noether current conservation reads 
\begin{equation}
\label{noethercurrent}
\nabla_{\mu} J^{\mu}=\frac{1}{\sqrt{-g}} \partial_{\mu}\left(\sqrt{-g} J^{\mu} \right)=0 \; , 
\end{equation}
where $J_{\mu} \equiv \frac{\delta S}{\delta\partial^\mu \varphi}=\lambda^2 \partial_{\mu} \varphi$ is the Noether current. The Noether charge density is given by
\begin{equation}
\label{noethercharge}
Q \equiv J_0=\lambda^2 \dot{\varphi} \; .
\end{equation}
As follows from Eq.~\eqref{noethercurrent}, it redshifts as $Q \propto \frac{1}{a^3}$ in the expanding Universe. So, during the inflationary expansion, the Noether charge density vanishes exponentially, and can be safely set to zero.  

In the case $Q=0$ there are two options. If the mass of field $\Psi$ is larger than the Hubble rate during inflation, its components $\Psi_1$ and $\Psi_2$ oscillate about their minima located at zero values and decay exponentially fast. 
No DM is generated in that case. In the opposite case of sufficiently light masses, $M \ll H$, components of the field $\Psi$ are in the slow roll regime during inflation. 
After inflation, when the Hubble parameter drops below the mass $M$, i.e., 
$H \lesssim M$, the complex scalar starts oscillating, and its further evolution is that of the pressureless perfect fluid. This scenario is standard, and we relegate further details to the Appendix. There we show, in particular, 
that the scenario with $M \ll H$ imposes severe constraints on the parameter space: mass $M$ itself and the scale of inflation. At least one of them 
should be very small in order to do not overproduce isocurvature perturbations.

It should be also stressed that no BA can be produced in the scenario considered in the present Section.
Indeed, BA is associated with the charge $Q$, which is zero. 
On the other hand, introducing $U(1)$-symmetry breaking interaction allows for generating the non-zero Noether charge $Q$ during inflation. 
We show this in the next Section.

\section{Generating Dark Matter during inflation}

\subsection{Quasi-de Sitter expansion of the Universe}

In the present Section, we discuss the complex scalar field coupled to
the inflaton. This, see below, opens up the opportunity of generating
super-heavy DM with  masses $M \gtrsim H$. Notably, the main
contribution to the relic abundance of DM can be produced during the
quasi-de Sitter expansion of the Universe. This is in constrast to the other mechanisms of the super-heavy DM production, 
which operate during preheating (see, e.g., the recent works~\cite{Gorbunov:2012ij, Kannike:2016jfs}). 

As in the bulk of the paper, we keep the ``DM reference''  for the complex scalar $\Psi$. However, let us reiterate that the inflaton is blind towards its nature, i.e., whether it carries baryon charge or not. 
Therefore, the mechanism for the production of the relic energy density of DM and BA, discussed below, is universal. There is, however, a difference in the strength of the coupling to the inflaton. 
We get back to this point, when relevant---in the end of Section 3.3.

We assume that the complex scalar field given by the action~(\ref{lambdadyn}) is coupled to the inflaton through an interaction term $S_{int}$, which explicitly breaks $U(1)$-symmetry. 
The simplest way to organize the symmetry breaking is by coupling the phase of the complex scalar 
to external matter. Note that coupling to the Standard Model species one risks to produce the large fifth force---in conflict with the 
Solar system tests\footnote{Normally one expects the heavy degrees of freedom to decouple in the limit $M \rightarrow \infty$. However, coupling to the matter fields through the phase is a special case, because the phase field is massless. Therefore, the decoupling in the large mass limit $M \rightarrow \infty$ may not take place.}. To be on the safe side, we assume that the phase interacts only with the inflaton. For the sake of concreteness, we choose the linear coupling of the phase to the 
trace of the inflaton stress-energy tensor $T_{infl} \equiv -\left(\partial_\mu\phi\right)^2 +4U(\phi)$,
\begin{equation}
\label{interaction}
S_{int}=\int d^4 x \sqrt{-g} \cdot  \beta \cdot \varphi \cdot T_{infl} \; ,
\end{equation}
where $\beta$ is dimensionless constant. Such a choice of coupling is not unique. 
Instead of the trace $T_{infl}$, one could assume a fairly arbitrary function of the inflaton field. 
The important things are: {\it (i)} it should remain almost constant
at inflation, {\it (ii)} the coupling breaks $U(1)$-symmetry, {\it
  (iii)} the coupling is linear in the dynamical field (phase
$\varphi$ in our case). This
leads to the production of the non-zero Noether charge as we will see
later. Note also that the interaction~(\ref{interaction}) is not
  periodic with respect to $\varphi$. Therefore, strictly speaking, 
  $\varphi$ cannot be a phase. However, one can think
  of~(\ref{interaction}) as an approximation of a periodic
  interaction, e.g. $\sim\sin(\beta\varphi)$ for small $\varphi$; in
  this case $\varphi$ is the phase.
  
With all the ingredients added up, the full action of our model reads,
  \begin{equation}
  \label{fullaction}
  	S = S_{EH} +S_\Psi + S_{int} + S_{infl}\,,
  \end{equation}
where $S_\Psi$, $S_{EH}$ and $S_{int}$ are given by Eqs.~\eqref{lambdadyn},~\eqref{actionEH} and~\eqref{interaction} correspondingly, while 
$S_{infl}$ is the standard action for the inflaton.

Varying action (\ref{fullaction}) with respect to $\varphi$, one gets in the homogeneous Universe,
\begin{equation}
\label{conserv}
\frac{1}{a^3} \frac{d}{dt} \left(Q a^3 \right)=\beta T_{infl} \; .
\end{equation}
Recall that $Q$ is defined by Eq.~\eqref{noethercharge}. In the slow roll regime, trace $T_{infl}$ is approximated by
\begin{equation}
\nonumber 
T_{infl} \approx 4U \; .
\end{equation}
Integrating Eq.~\eqref{conserv} we obtain,
\begin{equation}
\label{aboveequation}
Q (t) \approx \frac{4\beta}{a^3(t)} \int^{t}_{t_{in}} U(t') a^3 (t') dt' +\frac{C}{a^3(t)} \; .
\end{equation}
Here $C$ is the constant of integration referring to the pre-inflationary charge, which should be specified by the initial
conditions corresponding to the given class of inflationary
models. For instance, in case of the chaotic inflation all the relevant physical
quantities are of the Planckian order, while in case of the quantum
Universe production all of them (except for the energy density)
are zero. The related contribution (the last term in Eq.~\eqref{aboveequation}) 
quickly redshifts away during inflation, 
and one can safely neglect it. On the other hand, the first term on the r.h.s of Eq.~\eqref{aboveequation} does not disappear in the same manner, in spite of the factor $a^{-3}(t)$. To show it, note that in the inflationary Universe
the scale factor $a$ grows almost exponentially,
$a\propto \text{e}^{Ht}$, while $U$ and $H$ are slowly varying
  functions of time. 
Therefore, the integral in Eq.~\eqref{aboveequation} is saturated on the upper
boundary, and with the logarithmic accuracy one has 
\begin{equation}
\nonumber 
 \int^{t}_{t_{in}} U(t') a^3 (t') dt'\simeq \frac{a^3(t)U(t)}{3H(t)} \; .
\end{equation}
Finally, the estimate for the Noether charge density generated at some time $t$ during the quasi-de Sitter phase reads,
\begin{equation}
\label{interm}
Q  (t)\sim 4\beta \frac{U(t)}{3H(t)}\propto H \; .
\end{equation}
For comparison, in the exact de Sitter space-time approximation (with constant $U$ and $H$) one would obtain $Q (t)= \frac{4\beta U}{3H}$. This explains the factor '3' in the denominator of Eq.~\eqref{interm}. 
We see that the Noether charge density holds nearly constant, while its small variation with time is due to the non-zero slow roll parameters. 
 \begin{figure}[tb!]
\begin{center}
\includegraphics[width=0.8\columnwidth,angle=0]{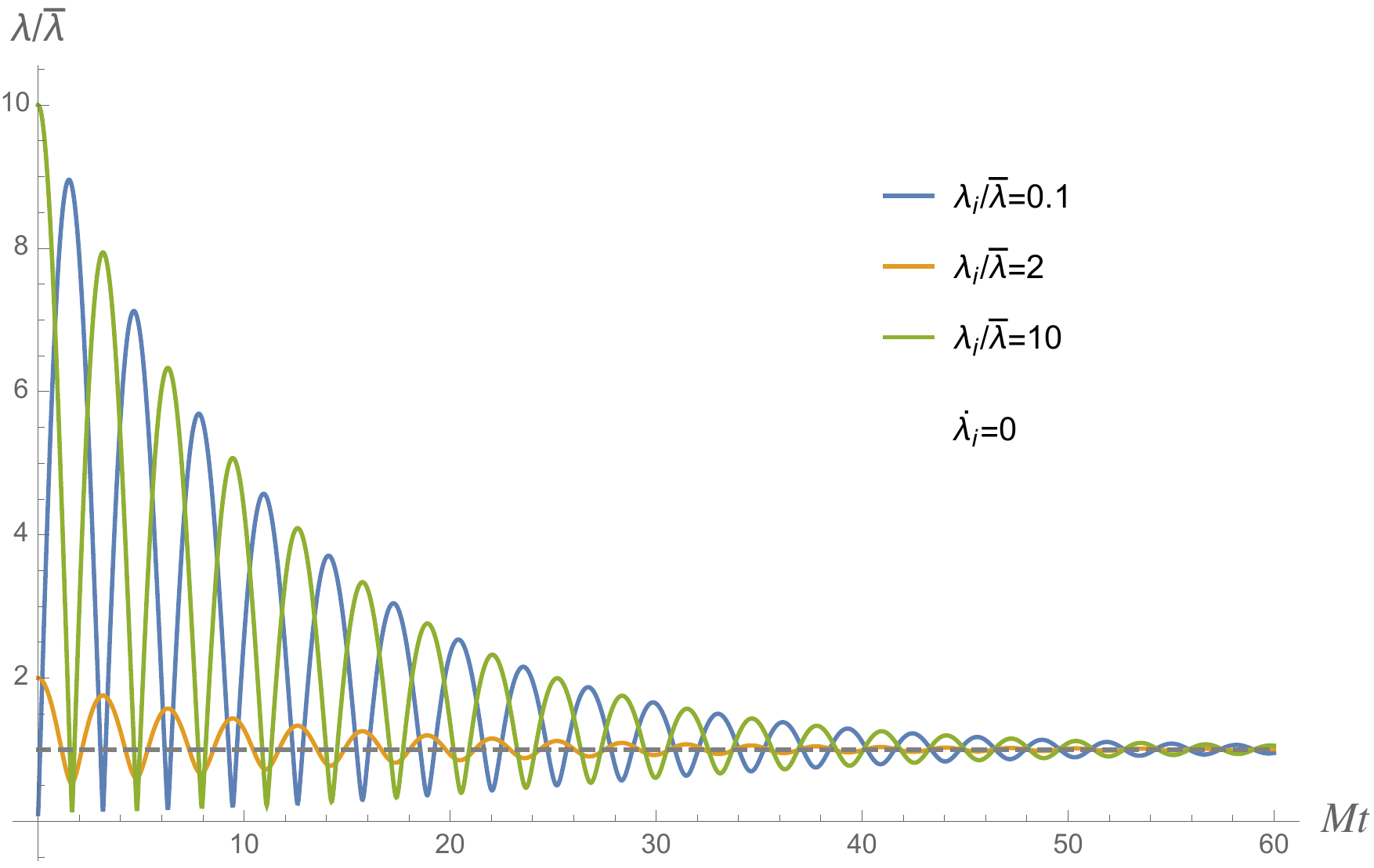}
\caption{Evolution of the field $\lambda$ is shown in the large mass regime $M \gtrsim H$, in the approximation of exact de Sitter space-time. We have chosen $\frac{M}{H}=20$. The initial value for the derivative $\dot{\lambda}$ 
is set to zero, $\dot{\lambda}_i=0$. The time range corresponds to 3 Hubble times. 
Independently of the initial value $\lambda_i$, the amplitude $\lambda$ relaxes to the minimum point $\bar{\lambda}$ (dashed line) of the effective potential~\eqref{effectivepotential}.}\label{largemass}
\end{center}
\end{figure}

Next we vary action~\eqref{fullaction} with respect to amplitude $\lambda$, 
\begin{equation}
\label{eqlambda} 
\square{\lambda}-\lambda \dot{\varphi}^2+M^2\lambda=0 \; .
\end{equation}
Expressing the time derivative of the field $\varphi$ from Eq.~\eqref{noethercharge}, we rewrite the latter equation as follows, 
\begin{equation}
\nonumber 
\ddot{\lambda}+3H \dot{\lambda}-\frac{Q^2}{\lambda^3}+M^2\lambda=0 \; .
\end{equation}
Hence, the field $\lambda$ resides in the effective potential 
\begin{equation}
\label{effectivepotential}
V_{eff}=\frac{M^2\lambda^2}{2}+\frac{Q^2}{2 \lambda^2} \; .
\end{equation} 
The minimum of potential \eqref{effectivepotential} is at $\bar{\lambda}$ given by
\begin{equation}
\label{lambdamin}
\bar{\lambda}=\sqrt{\frac{Q}{M}} \simeq \sqrt{\frac{4\beta U(t)}{3M H(t)}} \; ,
\end{equation}
where we made use of the estimate~\eqref{interm}.

So far our discussion has been independent of the mass $M$ of the complex field. Since this point on, however, this aspect becomes crucial. 
If the field $\Psi$ is substantially light, $M \ll H$, its amplitude
$\lambda$ enters the slow-roll regime at the onset of inflation akin to the
non-interacting (free) light complex scalar. This model has known problems
with isocurvature perturbations, as we mentioned in the Introduction and
illustrate with calculations in the Appendix. However, if 
inflation is sufficiently long, the Hubble parameter may decrease 
significantly, so that the field becomes heavy (compared to the Hubble scale) by the end of
inflation, $M > H$. One can expect such a scenario in chaotic
inflation with a power-law inflaton potential. The conclusion about large isocurvature perturbations is not applicable in that scenario. In the next Section, we show that they are exponentially suppressed compared to adiabatic ones. 
Thus, in the bulk of the paper, we mostly focus on the case 
of the super-heavy DM with the mass  
\begin{equation}
\nonumber 
M \gtrsim H \; ,
\end{equation} 
and hereafter by $H$ we assume somewhat loosely the value at $\sim 60$ e-folds 
before the end of inflation. 

 \begin{figure}[tb!]
\begin{center}
\includegraphics[width=0.8\columnwidth,angle=0]{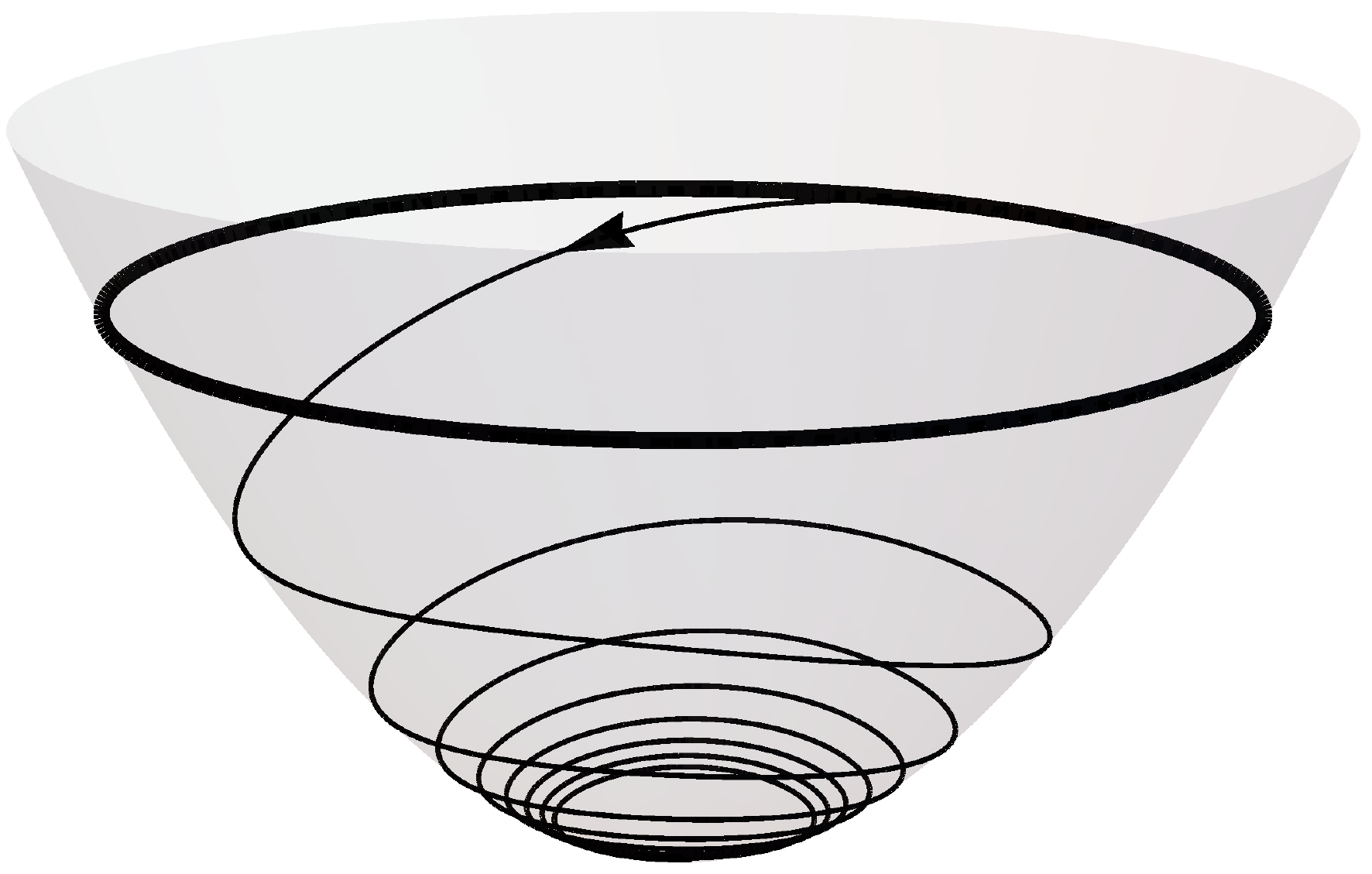}
\caption{Evolution of the complex scalar with the quadratic potential is shown during inflation and in the radiation-dominated epoch. The super-heavy DM with the mass $M \gtrsim H$ is considered. 
During inflation, the amplitude of the complex field resides in the minimum of the effective potential~\eqref{effectivepotential}, $\lambda= \bar{\lambda}$, where $\bar{\lambda}= \sqrt{\frac{Q}{M}}$. The amplitude is nearly constant in the (quasi) de Sitter space-time. Consequently, 
the trajectory of the complex field in the configuration space is almost circular. The uncertainties about the circle (captured by the thick line) reflect small deviations from the de Sitter space-time due to non-zero 
slow roll parameters. After the end of inflation, the trajectory is inspiralling 
with the radius decreasing as $1/a^{3/2}$. The latter reflects the drop of the amplitude $\lambda$ in the hot Universe.}\label{vedro}
\end{center}
\end{figure}

In Fig.~\ref{largemass}, we show the evolution of the 
amplitude $\lambda$. We see that it relaxes to the minimum of the effective potential $\bar{\lambda}$ within a few Hubble times independently on its initial conditions. That is, the solution~\eqref{lambdamin} is an attractor. We simplified 
the picture by considering the exact de Sitter space-time, in which case the solution~\eqref{lambdamin} is just a constant. In reality, however, the field $\lambda$ does not get frozen out completely. 
Indeed, the parameters of the effective potential~\eqref{effectivepotential} are changing with time---though, slowly. Therefore, the value of $\lambda$ corresponding to the minimum of the effective potential also shifts with time, 
\begin{equation}
\label{dotlambdamin}
\dot{\bar{\lambda}} \sim M_{Pl} \dot{H}\sqrt{\frac{\beta}{M H}} \; .
\end{equation} 
The fact that $\dot{\bar{\lambda}}$ is non-zero is relevant for the
evolution of the adiabatic perturbation mode discussed in Section~4.

While the field $\lambda$ takes a nearly constant value, the phase ``velocity'' also approaches constant $\dot{\varphi} \rightarrow M$. This follows from Eq.~\eqref{eqlambda}, where one 
sets time derivatives of the amplitude $\lambda$ to zero. We conclude that during the inflationary stage, the complex scalar spins fast in the configuration space with nearly constant amplitude and frequency, see Fig.~\ref{vedro}.

\subsection{Preheating and later stages}

After the end of inflation, the force associated with the inflaton decreases rapidly, and its further impact on the Noether charge
evolution vanishes. Since this point on, the Noether charge density drops with the scale factor as $Q \sim 1/a^3$, i.e.,  
\begin{equation}
\label{chargeredshift}
Q (t>t_{end})\simeq Q(t=t_{end})\frac{a^3_{end}}{a^3} \simeq
 \frac{4\beta}{a^3(t)}\frac{a^3 (t_{end})U(t_{end})}{3H(t_{end})}  \; .
\end{equation}
Here $t_{end}$ designates the end of inflation, i.e., the moment,
when slow roll parameters $\epsilon$ and $\eta$ become of the order
unity, i.e., $\epsilon \sim 1$ and $\eta \sim 1$. That estimate holds,
e.g., 
if we assume the immediate decay of the inflaton to the particle species. In reality, the evolution of the inflaton during the preheating stage may considerably alter this result. That is, the actual integral, which gives the contribution to the total 
Noether charge density is given by, 
\begin{equation}
\label{integralsplit}
\int^{t_{end}}_{t_{in}} \rightarrow \int^{t_{end}}_{t_{in}}+ \int^{t_{preheating}}_{t_{end}} \; ,
\end{equation}
where $t_{preheating}$ denotes the end of preheating. Generically, the contribution due to the second integral on the r.h.s. is of the order of the first one (if preheating continues for a few Hubble times) or much larger than it (if preheating is long enough in terms of Hubble times). 
However, there are situations, which we are interested in, when the main contribution comes from the quasi-de Sitter 
expansion of the Universe. Apart from the general class of inflationary
models with instant preheating, this occurs in a number of
specific models. For example, consider the non-minimally coupled
inflaton with the quartic potential~\cite{Salopek:1988qh,
  Futamase:1987ua, Okada:2010jf, Linde:2011nh, Bezrukov:2013fca}, 
\begin{equation}
\nonumber 
S_{infl}=\int d^4 x \sqrt{-g} \left(\frac{1}{2} (\partial \phi )^2+\frac{\xi}{2} R \phi^2-\frac{h}{4} \phi^4 \right) \; .
\end{equation}
The role of the non-minimal coupling is to reduce the tensor-to-scalar ratio, which would otherwise violate the existing Planck bound~\cite{Ade:2015lrj}. 
The trace of the stress-energy tensor $T_{infl}$ rapidly decreases after the end of inflation following the restoration of the conformal 
symmetry in the cosmological background. Hence in this case, no DM is generated during the preheating epoch.

After the inflaton decays, the complex scalar $\Psi$ becomes free modulo interactions with Standard Model species mediated by the virtual inflaton. 
Given that $M \gtrsim H$, its components start oscillating around the origin with the amplitude decaying as $1/a^{3/2}$. Since this point on, the complex scalar mimics the pressureless perfect fluid: its energy density redshifts as 
$\rho_{DM} (t) = M^2 \lambda^2 (t) \propto 1/a^3(t)$. At the onset of oscillations, which start right after the end of inflation at the time $t=t_{end}$, 
the amplitude $\lambda$ is estimated by Eq.~\eqref{lambdamin}, and hence the DM energy density is given by 
\begin{equation}
\label{dmosc}
\rho_{DM} (t_{end}) \simeq M^2 \bar{\lambda}^2 \simeq M Q (t_{end}) \; .
\end{equation}
We see that it is fully determined by the Noether charge density generated and the mass of the scalar, 
and is {\it independent of the pre-inflationary value of the field $\Psi$.} Then it decreases following the behavior of the charge density $Q(t)$ as is written in Eq.~\eqref{chargeredshift}. We require that the energy density~\eqref{dmosc} should match the abundance of DM observed today. Concretely, 
\begin{equation}
\rho_{DM} (t_{eq}) \simeq \frac{\beta M U(t_{end})}{H(t_{end})} \cdot \frac{a^3_{{end}}}{a^3_{eq}}  \simeq \rho_{rad} (t_{eq}) \; ,
\end{equation}
where the subscript $'eq'$ denotes the equality between the matter and the radiation. For the sake of simplicity, we assume that the radiation is produced immediately at the very end of inflation, so that $\rho_{rad} (t_{end}) \simeq U(t_{end})$. 
Then one can write 
\begin{equation}
\nonumber 
\rho_{rad} (t_{eq}) \simeq \rho_{rad} (t_{end}) \cdot \frac{a^4_{end}}{a^4_{eq}} \simeq U(t_{end}) \cdot \frac{a^{4}_{end}}{a^4_{eq}} \; . 
\end{equation}
(In the first estimate here we neglected the change of the number of
ultra-relativistic degrees of freedom during radiation-dominated
epoch, which is straightforward to account for.)  The estimate for the constant $\beta$ then takes the form, 
\begin{equation}
\beta \simeq \frac{H(t_{end})}{M} \cdot \frac{a_{end}}{a_{eq}} \simeq \frac{H(t_{end})}{M} \cdot \frac{T_{eq}}{T_{end}} \; .
\end{equation}
Substituting the estimate for the Hubble rate $H(t_{end}) \sim \frac{T^2_{end}}{M_{Pl}}$, we obtain for the coupling constant $\beta$, 
\begin{equation}
\beta \simeq \frac{T_{end}}{M_{Pl}} \cdot \frac{T_{eq}}{M}  \; .
\end{equation}
Taking for instance $T_{end} \sim 10^{16}$~GeV (maximal possible
temperature in the post-inflationary Universe) and $M \sim H \sim
10^{-5} M_{Pl}$, one gets $\beta \sim 10^{-26}$. This is a rough estimate, which first assumes high scale inflation and then does not
account for the details of the cosmological evolution during 
(p)reheating. However, the message is clear: production of super-heavy
complex field DM requires a tiny coupling to the inflaton. Indeed, we produce non-relativistic relics even before the 
radiation-dominated stage, where their relative contribution to the
Universe energy density grows linearly with the scale factor. In the case of
almost free relics, the
growth must be compensated and only the strong suppression of the
production rate can help. This is achieved by choosing a small value of the coupling constant. Since
the coupling violates global $U(1)$-symmetry, it might be attributed
to some non-perturbative quantum effect, so that the smallness may be
natural. 

\begin{figure}[tb!]
\begin{center}
\includegraphics[width=0.8\columnwidth,angle=0]{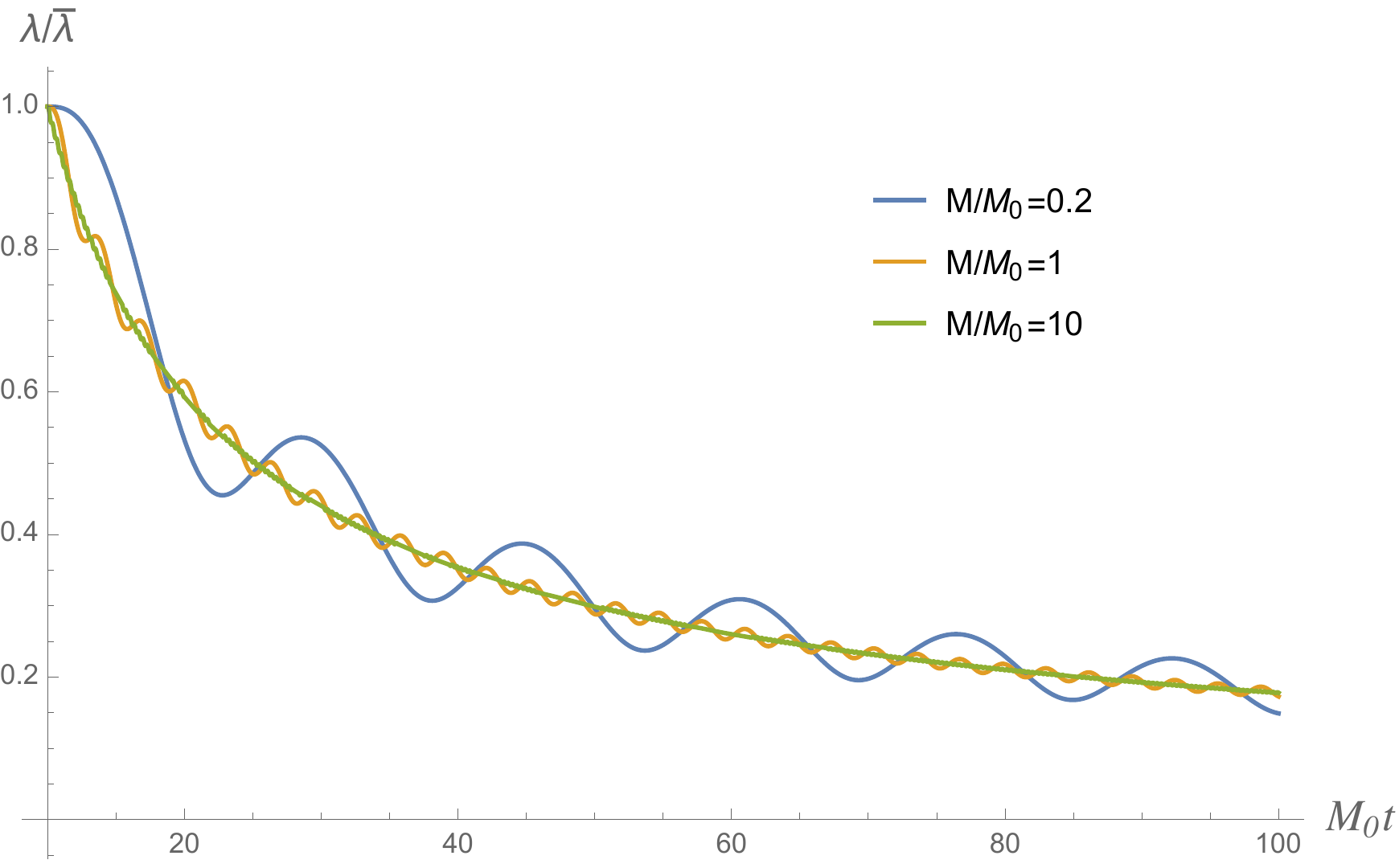}
\caption{Evolution of the field $\lambda$ is shown in the large mass regime, i.e., $M \gg H$, during the radiation-dominated stage. We assume for simplicity that the latter follows immediately after inflation. 
The initial conditions at the beginning of the hot epoch (coinciding with the end of inflation, time $t_{end}$) are $\lambda (t_{end})=\bar{\lambda}$ and $\dot{\lambda} (t_{end})=0$. The characteristic mass scale $M_0$ is chosen to be such that $\frac{M_0}{H(t_{end})} =20$. For the masses $M \lesssim M_0$, the evolution 
of the amplitude exhibits clear oscillations about the average
value. Instead, for $M\gg M_0$ oscillations are very small, and the
change of the field $\lambda$ tracks the cosmological expansion, i.e.,
$\lambda \propto \frac{1}{a^{3/2}}$. In that case  
the evolution of the amplitude can be inferred from the simpler model~\eqref{mimetic}, where $\lambda$ plays the role of the Lagrange multiplier.}\label{mim}
\end{center}
\end{figure}

Let us make one remark in passing. In the limit, when the field $\lambda$ is initially frozen and $M \rightarrow \infty$, the model of the complex scalar effectively reduces to the following one, 
\begin{equation}
\label{mimetic0} 
S=\int d^4 x \sqrt{-g} \left[\frac{\lambda^2}{2} (\partial_{\mu} \varphi)^2-\frac{M^2\lambda^2}{2} \right] \; ,
\end{equation}
see Fig.~\ref{mim}. 
Namely, the dynamics of the amplitude $\lambda$ freezes out, and the latter reduces to the Lagrange multiplier. Upon redefining the amplitude and the phase, i.e., $\lambda \rightarrow \lambda /M$ and $\varphi \rightarrow M \varphi$, 
the action~\eqref{mimetic0} takes the form 
\begin{equation}
\label{mimetic} 
S=\int d^4x \sqrt{-g} \cdot \frac{\lambda^2}{2} \cdot \left[(\partial_{\mu} \varphi)^2-1 \right] \; .
\end{equation}
The stress-energy tensor of the model \eqref{mimetic} is $T_{\mu \nu}=\lambda^2 \partial_{\mu} \varphi \partial_{\nu} \varphi$. 
It corresponds to the pressureless perfect fluid with $\lambda^2$
being the energy density and $\varphi$ the velocity potential. We see
that at $M\to\infty$  the complex scalar exactly reproduces the pressureless perfect fluid. 
(Normally the pressureless perfect fluid is obtained from the dynamics of the scalar field upon averaging 
over its oscillations.)
This holds true down to the times, when caustics are formed and the fluid description breaks down. On the other hand, the model of the complex field always retains a non-singular behavior.  

The model~\eqref{mimetic} and its variations is nowadays widely discussed in the literature 
and goes under the name of the mimetic matter scenario~\cite{Chamseddine:2013kea, Chamseddine:2014vna}\footnote{The original formulation of the mimetic matter scenario is different from Eq.~\eqref{mimetic}. It stems from the particular conformal 
transformation of the metric. The equivalence with the model~\eqref{mimetic} has been first pointed out in Ref.~\cite{Golovnev:2013jxa}.}. Originally, it has been designed as an alternative to the particle DM. According to our discussion above, the model~\eqref{mimetic} breaks down at the times of the caustic formation and must be completed. The super-heavy complex scalar with initial conditions set by inflation may serve as a proper candidate for this completion~\cite{Babichev:2017lrx}.

\subsection{Baryon asymmetry}

While we mainly discussed DM so far, in an alternative scenario the generated Noether charge
can explain BA of the Universe. For this scenario to be realized, one assigns the non-zero baryon charge to the complex field $\Psi$. 
The rest of the scenario can be designed by analogy with the standard mechanisms of BA generation: the baryon charge should be transferred from 
$\Psi$ to the quark sector of the Standard Model. Assume for concreteness that the baryon charge of the field $\Psi$ equals unity, $B_{\Psi}=1$. Then, one can write down the following interaction, 
\begin{equation}
\nonumber 
{\cal L} =y \bar{n}S \Psi+h.c. \; ,
\end{equation}  
where $y$ is the coupling constant, $n$ is the colorless fermionic operator formed by three Standard Model quarks (e.g., neutron-like fermion, $n=ddu$) and $S$
is another fermionic field, which is a singlet with respect to the Standard Model gauge group (e.g., sterile neutrino). Then BA in the quark sector is induced through the decays
of the field $\Psi$.
The resulting BA is given by 
\begin{equation}
\nonumber
\Delta_B \sim \frac{Q}{s} \; ,
\end{equation} 
where $s$ is the entropy density.

Compared to the DM case, production of BA generically requires much larger coupling constant $\beta$. Let us show this explicitly. For this purpose, we again assume that the inflaton energy 
is converted immediately to radiation right after inflation. Then, 
\begin{equation}
\nonumber 
\Delta_B \simeq \frac{\beta U (t_{end}) T_{end}}{H (t_{end}) \rho_{rad} (t_{end})} \simeq \frac{\beta T_{end}}{H (t_{end})} \simeq \beta \cdot \sqrt{\frac{M_{Pl}}{H (t_{end})}}\; .
\end{equation}
Here we use that $s = \frac{4\rho_{rad}}{3T}$ and $T \simeq \sqrt{M_{Pl} H}$. The estimate for the coupling constant then follows from the observed BA of the Universe, $\Delta_B =0.87 \cdot  10^{-10}$. We have 
\begin{equation}
\label{estbetaba}
\beta \simeq 10^{-10} \sqrt{\frac{H (t_{end})}{M_{Pl}}} \; .
\end{equation}
Hence, for the high scale inflation with the Hubble rate $H \simeq 10^{14}$~GeV, the coupling constant $\beta$ can be as large as $\beta \simeq 10^{-12}$.

Some remarks are in order here. The estimate~\eqref{estbetaba} does not depend on the lifetime of 
the field $\Psi$, since both $Q$ and $s$ drop similarly with the scale factor as the Universe enters the radiation-domination regime. On the other hand, this estimate
changes if the energy density of the Universe scales differently than radiation after inflation. Say, for the matter-domination stage at preheating
the relevant amount of BA requires bigger coupling by a
ratio of the scale factors at reheating and at the end of
inflation. 

As we can see, scenarios for DM and BA generation involve coupling constants $\beta$ different by the factor $\sim 10^{-14}$.
The appearance of this factor can be understood as follows. Assuming instant preheating the ratio of the
coupling constants required for DM and BA generation is about the ratio of the proton mass (the lightest stable
baryon) and the Hubble parameter at inflation (the smallest mass of
the viable heavy DM), i.e., $\sim m_p/H\sim10^{-14}$.

\section{Perturbations}

Now let us discuss perturbations of the complex scalar field. We want to prove that only adiabatic fluctuations survive in the case of the super-heavy DM, while isocurvature modes quickly decay behind the 
horizon. This is relevant in view of the strong Planck limits on the Cold DM isocurvature 
perturbations~\cite{Ade:2015lrj}. On the other hand, for lighter DM, there is a non-decaying isocurvature mode, which can be large enough, unless the scale of inflation/the mass of DM are strongly constrained. 
This case is considered in the Appendix.

We choose to study the linear perturbations in the Newtonian gauge~\cite{Gorbunov:2011zzc, Mukhanov:1990me}, 
\begin{equation}
\nonumber 
ds^2=(1+2\Phi)dt^2-a^2(1-2\Psi) \delta_{mn}dx^m dx^n \; ,
\end{equation}
so that 
\begin{equation}
\nonumber 
\delta g_{00}=2\Phi\,, \qquad \delta g_{ij}=2a^2 \Psi \delta_{ij} \; .
\end{equation}
From $i\neq j$ components of the perturbed Einstein equations we have $\Psi=\Phi$, and the $0i$-component takes the form
\begin{equation}
\label{eompotential}
\dot{\Phi}+H\Phi=4\pi G \left(\dot{\phi} \delta \phi +\dot{\lambda} \delta \lambda +\lambda^2 \dot{\varphi} \delta \varphi \right) \; .
\end{equation}
The linearized equations of motion for the amplitude $\lambda$ and the phase $\varphi$ are given by
\begin{equation}
\label{eomlambda}
\delta \ddot{\lambda}+3H \delta \dot{\lambda} -\frac{1}{a^2} \partial_i \partial_i \delta \lambda -\delta \lambda \dot{\varphi}^2 -2\lambda \dot{\varphi} \delta \dot{\varphi} +
M^2 \delta \lambda +2M^2 \lambda \Phi -4 \dot{\lambda} \dot{\Phi}=0 
\end{equation}
and
\begin{equation}
\label{eomphase}
\begin{split}
\frac{1}{a^3} \frac{d}{dt}\Bigl(a^3 \lambda^2 \delta \dot{\varphi} \Bigr) -\frac{1}{a^2}\lambda^2  \partial_i \partial_i \delta \varphi  -\frac{2}{a^3} \frac{d}{dt} \Bigl(a^3 \lambda^2 \dot{\varphi} \Bigr) \Phi 
-4\lambda^2 \dot{\varphi} \dot{\Phi}+\frac{1}{a^3}\frac{d}{dt} \Bigl(a^3 \dot{\varphi} \delta \lambda^2 \Bigr)=\beta \delta T_{infl} \; ,
\end{split}
\end{equation}
respectively.

The analysis of perturbations parallels to that of inflation with multiple scalar fields. The adiabatic mode is given by~\cite{Polarski:1994rz}
\begin{equation}
\label{ansatzpotential} 
\Phi=\Psi=C_1 \left(1-\frac{H}{a} \int^{t}_0 a dt' \right)+C_2 \frac{H}{a} 
\end{equation}
and
\begin{equation}
\label{ansatzfields}
\frac{\delta \phi}{\dot{\phi}}=\frac{\delta \lambda_{ad}}{\dot{\lambda}}=\frac{\delta \varphi_{ad}}{\dot{\varphi}}=\frac{1}{a} \left(C_1 \int^t_0 a dt' -C_2 \right) \; .
\end{equation}
Here $C_1$ and $C_2$ are constants defined from the matching to the inflaton vacuum initial conditions. They correspond to the growing and decaying adiabatic modes, respectively. 
One can check that in the super-horizon regime the ansatz~\eqref{ansatzpotential} and~\eqref{ansatzfields} passes the equations of motion~\eqref{eompotential},~\eqref{eomlambda} and~\eqref{eomphase}. 
Note that the solution~\eqref{ansatzpotential} and~\eqref{ansatzfields} is exact. The equalities~\eqref{ansatzfields} can be easily reformulated in terms of the 
complex scalar components $\Psi_1$ and $\Psi_2$, 
\begin{equation}
\nonumber 
\frac{\delta \phi}{\dot{\phi}}=\frac{\delta \Psi_{1,ad}}{\dot{\Psi}_1}=\frac{\delta \Psi_{2,ad}}{\dot{\Psi}_2} \; .
\end{equation}
Notably, despite the non-trivial coupling to the inflaton via the phase field, the adiabatic condition has the same form as in the case of 
canonical non-interacting fields. 

Let us switch to the case of isocurvature perturbations. For this purpose, one sets to zero gravitational potentials, 
\begin{equation}
\nonumber 
\Phi=\Psi=0 \; .
\end{equation} 
In the spectator approximation, which we assume here, the inflaton is the only source of metric fluctuations. 
Therefore, one may consistently set the isocurvature perturbation 
of the inflaton to zero, $\delta \phi_{iso}=0$. Then the
relevant system of equations is given by
\begin{equation}
\label{systemone}
\delta \ddot{\lambda}_{iso}+3H \delta \dot{\lambda}_{iso} -\frac{1}{a^2} \partial_i \partial_i \delta \lambda_{iso} -\delta \lambda_{iso} \dot{\varphi}^2 -2\lambda \dot{\varphi} \delta \dot{\varphi}_{iso} +
M^2 \delta \lambda_{iso} =0 
\end{equation}
and
\begin{equation}
\label{systemoneone}
\begin{split}
\frac{1}{a^3} \frac{d}{dt}\Bigl(a^3 \lambda^2 \delta \dot{\varphi}_{iso} \Bigr) -\frac{1}{a^2}\lambda^2  \partial_i \partial_i \delta \varphi_{iso}  +\frac{1}{a^3}\frac{d}{dt} \Bigl(a^3 \dot{\varphi} \delta \lambda^2_{iso} \Bigr)=0\; .
\end{split}
\end{equation}
In the super-horizon regime, Eq.~\eqref{systemoneone} takes the form 
\begin{equation}
\nonumber 
\frac{d}{dt} \left(a^3 \delta [\lambda^2 \dot{\varphi}] _{iso}\right)=0 \; .
\end{equation}
Consequently, one has 
\begin{equation}
\nonumber 
\delta (\lambda^2 \dot{\varphi})_{iso}=\frac{C}{a^3} \; .
\end{equation}
Hence, the isocurvature perturbation of the product $\lambda^2 \dot{\varphi}$ decays fast behind the horizon, and one can write
\begin{equation} 
\label{corr}
\frac{\delta \dot{\varphi}_{iso}}{\dot{\varphi}}=-\frac{2\delta \lambda_{iso}}{\lambda} \; .
\end{equation}
Expressing the perturbation $\delta \dot{\varphi}$ from the latter and substituting into Eq.~\eqref{systemone}, we obtain 
\begin{equation}
\nonumber 
\delta \ddot{\lambda}_{iso}+3H \delta \dot{\lambda}_{iso}+M^2_{eff} \delta \lambda_{iso}=0 \; ,
\end{equation} 
where 
\begin{equation}
\label{effectivemass} 
M^2_{eff}=M^2+3\dot{\varphi}^2 \; .
\end{equation}
As we assume the super-heavy DM with the mass $M \gtrsim H$, one has
$M_{eff} \gtrsim H$.
Hence, the perturbation $\delta \lambda_{iso}$ relaxes to zero within a few Hubble times, 
independently of its initial value. At the same time, the adiabatic perturbation $\delta \lambda_{ad}$ remains constant behind the horizon, 
\begin{equation}
\nonumber 
\delta \lambda_{ad} \sim \frac{\dot{\lambda}}{\dot{\phi}} \delta \phi \sim \sqrt{\frac{\beta H|\dot{H}|}{M}} \sim \sqrt{\frac{\epsilon \beta H^3}{M}}\; .
\end{equation}
Here we use that $\dot{\phi} \sim \sqrt{M^2_{Pl} |\dot{H}|}$, $\delta \phi \sim H$ and substitute Eq.~\eqref{lambdamin}. We conclude that for $M \gtrsim H$ perturbations are adiabatic with a very high accuracy.

On the contrary, for the light DM ($M \ll H$) discussed in the Appendix, the fate of perturbations is not that different from the case of the non-interacting complex field. Namely, isocurvature perturbations dominate over adiabatic ones, 
unless the scale of inflation is low and/or the mass $M$ is small. This 
severely constrains the scenario of light DM. Nevertheless, slightly extending the parameter space 
of the model allows one to suppress the isocurvature mode for essentially
arbitrary $M$. We consider such a modification in the next Section.

\section{Non-minimal coupling to gravity}

\subsection{Inflationary stage}

Having super-heavy complex field is not the only possibility to suppress isocurvature modes. 
Another option, which we consider in the present Section, is to couple the amplitude of the complex scalar 
to the Ricci curvature. Similar ideas have been proposed in the context of axion models in Refs.~\cite{Linde:1991km, Dine:2004cq}. 
We will see that in the presence of the non-minimal coupling, the mass of the field $\Psi$ remains essentially unconstrained. Such a freedom 
is not particularly relevant in the scenario, where the complex scalar carries baryon charge. On the other hand, it is important for the 
DM applications of our model. In particular, if the latter is ultra-light, the composition of DM halos may have distinct features compared to the standard Cold DM. 
Furthermore, for light DM, the coupling constant $\beta$ is allowed to
be much larger than in the case of super-heavy DM.

We modify the model by introducing the non-minimal coupling of $\Psi$ to gravity,
\begin{equation}
\nonumber 
S_{non-min}=\int d^4 x \sqrt{-g} \left(\frac{\xi \lambda^2}{2} R  \right) \; .
\end{equation}
During the inflationary expansion, the Ricci scalar $R=-12H^2-6\dot{H}$ holds nearly constant. Thus, the non-minimal coupling mimics the mass term of the complex field. 
The equation of motion~\eqref{eqlambda} for the field $\lambda$ now takes the form, 
\begin{equation}
\nonumber 
\ddot{\lambda}-\lambda \dot{\varphi}^2+3H \dot{\lambda}+M^2_{eff} \lambda=0 \; ,
\end{equation}
where the effective mass squared $M^2_{eff}$ is the sum of two contributions: one coming from the standard mass term and another due to the non-minimal coupling, i.e., 
\begin{equation}
M^2_{eff}\approx M^2+12\xi H^2 \; .
\end{equation}
(we made use of the de Sitter approximation at this step). Note that the non-minimal coupling is irrelevant for the generation of the Noether charge density (as it does not break the $U(1)$-invariance). 
That is, Eq.~\eqref{noethercharge} holds true, and one can use it to express the time derivative $\dot{\varphi}$. Repeating the steps from Section 3.1, we find the effective potential for the field $\lambda$, 
\begin{equation}
\label{effectivepotentialnew}
V_{eff}=\frac{M^2_{eff} \lambda^2}{2}+\frac{Q^2}{2\lambda^2} \; .
\end{equation}
It coincides with the standard effective potential~\eqref{effectivepotential} modulo the replacement $M \rightarrow M_{eff}$.
Thus, the minimum point of the new effective potential is given by Eq.~\eqref{lambdamin}, where one should just substitute $M_{eff}$ instead of $M$. If the effects 
of the non-minimal coupling are weak, so that $M_{eff} \approx M$, we return to the situation described in the previous Sections. Here we are interested in the low mass limit, i.e., 
\begin{equation}
\nonumber 
M \ll \sqrt{\xi} H \; .
\end{equation}
In that case, the non-minimal coupling dominates over the standard mass term, so that $M_{eff} \sim \sqrt{\xi} H$. Substituting the latter into Eq.~\eqref{lambdamin}, one gets for the minimum point 
of the new effective potential~\eqref{effectivepotentialnew},
\begin{equation}
\label{lambdaminnonm}
\bar{\lambda}_{non-min} \simeq \frac{\sqrt{\beta} M_{Pl}}{{\xi}^{1/4}} \; .
\end{equation}
(The subscript 'non-min' is added to distinguish this minimum point from the one obtained in the case of super-heavy DM, see Eq.~\eqref{lambdamin}.)
During inflation, the amplitude $\lambda$ relaxes to $\bar\lambda$ within a few Hubble times, as in the case of the 
genuinely super-heavy complex field. See Fig.~\ref{largemass}.

 \begin{figure}[tb!]
\begin{center}
\includegraphics[width=0.8\columnwidth,angle=0]{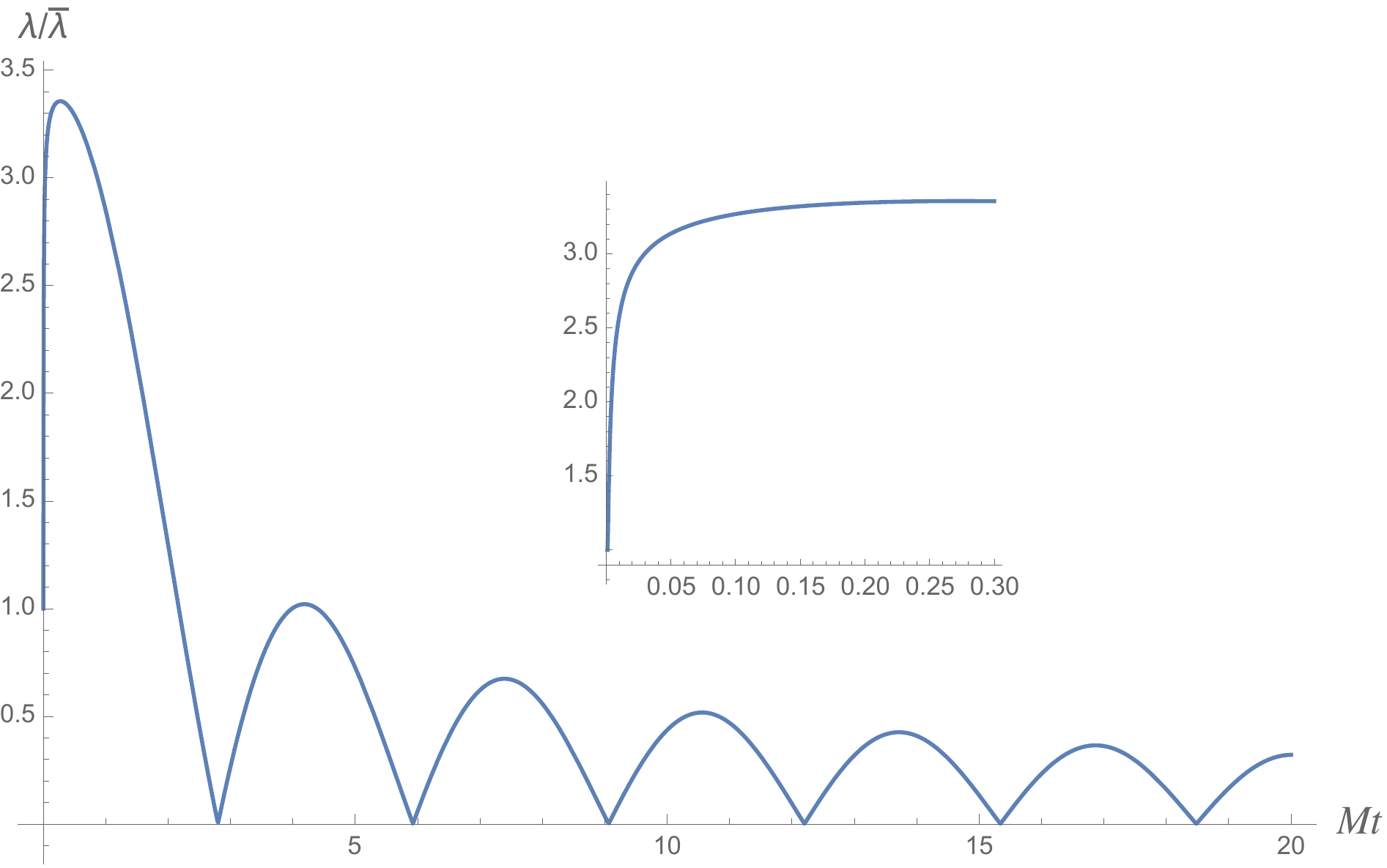}
\caption{Evolution of the field $\lambda$ is shown during the radiation-dominated stage in the version of the model with the non-minimal coupling to gravity. We have chosen the coupling to the Ricci curvature $\xi=1$. The small mass $M$ is assumed, i.e., $\frac{M}{H (t_{end})}=2 \cdot 10^{-3}$ and 
$Mt_{end}=10^{-3}$. At the times $Mt \gtrsim 1$, the 
complex field exhibits damping oscillations around the minimum. We zoomed the non-trivial part of the evolution taking place at early times $Mt \ll 1$. Initially the amplitude $\lambda$ rolls fast; this reflects the instantaneous change 
of the structure of the effective potential~\eqref{effectivepotential} after inflation. Then the evolution of the field $\lambda$ slows down for the reasons described in the text. At the times $Mt \sim 1$ the oscillations start and the complex field enters the dust-like phase.}\label{nonminimal}
\end{center}
\end{figure}

The fate of adiabatic and isocurvature perturbations is also similar. The former obey Eqs.~\eqref{ansatzpotential} and~\eqref{ansatzfields} {\it in the Einstein frame}, while the latter rapidly decay behind the horizon provided only that the parameter $\xi$ governing 
the non-minimal coupling is not small, i.e., $\xi \gtrsim 1$. Thus, with the non-minimal coupling to gravity, the scenario with {\it light} DM may be viable as well.

\subsection{After inflation}

 We assume for the sake of simplicity that the radiation-dominated stage follows immediately after the end of the quasi-de Sitter expansion. 
Then, the Ricci scalar $R$ abruptly drops to zero, $R \rightarrow 0$, and the mass term of the complex scalar 
takes the standard form. This leads to the abrupt change of the effective potential~\eqref{effectivepotentialnew}, so that we recover our original expression~\eqref{effectivepotential},
\begin{equation}
\nonumber 
V_{eff}=-\frac{\xi R \lambda^2}{2}+\frac{Q^2}{2\lambda^2} \rightarrow \frac{M^2\lambda^2}{2}+\frac{Q^2}{2\lambda^2} \; .
\end{equation}
(we assumed that $M^2 \ll |\xi R|$ during inflation). Now, the minimum of the  effective potential is at $\bar{\lambda}$ given by Eq.~\eqref{lambdamin}. Note that $\bar{\lambda} \gg \bar{\lambda}_{non-min}$. Consequently, the amplitude $\lambda$, which still resides at $ \lambda =\bar{\lambda}_{non-min}$ right after inflation, is away from the location of the true minimum $\bar{\lambda}$. Then, right after the end of inflation the amplitude $\lambda$ rolls towards $\bar{\lambda}$. This is reflected in the initial fast growth of 
$\lambda$ shown in Fig.~\ref{nonminimal}.

Further evolution of the amplitude proceeds as follows. The complex scalar starts with a large kinetic energy due to rotations induced by the coupling to the inflaton, while the potential energy $\sim M^2 \lambda^2$ is yet negligible. 
At those times, the derivative of the amplitude decreases fast, $\dot{\lambda} \sim \frac{1}{a^3}$, and hence the growth of the amplitude quickly stabilizes.
Shortly, the kinetic energy, which redshifts as $ \sim \frac{1}{a^6}$, drops below the potential energy, which is kept nearly constant, and the slow roll phase starts. At later times $t \sim H^{-1} \gtrsim M^{-1}$, the field $\Psi$ exhibits the standard oscillatory behavior, and its energy density $\rho_{DM} \sim M^2 \lambda^2 \sim 1/a^3$ mimics that of the pressureless perfect fluid. 

As a result, the amplitude $\lambda$ does not change considerably by the beginning of oscillations and can be roughly estimated by Eq.~\eqref{lambdaminnonm}. Hence, its value (and, thus, the relic energy density of DM) is attached to the Noether charge density generated by the end of inflation, 
as in the case of the super-heavy complex scalar. The combination of constants $\beta$, $\xi$ and the mass $M$ can be constrained from the observed relic abundance of DM. Namely, at the matter-radiation equality, DM energy density is estimated by,
\begin{equation}
\nonumber 
\rho_{DM} (t_{eq})\sim M^2 \lambda^2 (t_{eq})\sim M^2 \bar{\lambda}^2_{non-min} \cdot  \frac{T^3_{eq}}{T^3_{H \sim M}} \sim \rho_{rad} (t_{eq}) \sim T^4_{eq}\; ,
\end{equation} 
where $T_{eq}$ is the equilibrium temperature, $T_{eq} \sim 1$~eV and $T_{H \sim M}$ is temperature of the Universe at the onset of oscillations, $T_{H \sim M} \sim \sqrt{M M_{Pl}}$. 
We get an estimate of the coupling constant $\beta$,
\begin{equation}
\nonumber 
\beta \sim \frac{\sqrt{\xi}}{\sqrt{M \cdot M_{Pl}}} \cdot T_{eq} \; .
\end{equation}
For example, the widely discussed in literature ultra-light scalar field DM of mass $M \simeq 10^{-21}$~eV requires coupling $\beta \sim 10^{-3}$. 
Thus, in the scenario with the non-minimal coupling to gravity, the
constant $\beta$ is not necessarily very small. This is in contrast to the case of super-heavy DM considered in the previous Sections.

The field DM with such low masses may affect the standard picture of the halo formation.
Generically, classical field DM obeys the Schr$\ddot{\mbox{o}}$edinger--Poisson system of 
equations~\cite{Widrow:1993qq, Uhlemann:2014npa, Kopp:2017hbb, Mocz:2018ium}, 
while the standard Cold DM obeys the Vlasov--Poisson system. The difference between two scenarios is particularly strong in the ultra-light case~\cite{Hu:2000ke, Woo:2008nn, Hui:2016ltb}. This has been used to set the lower limit on the axion mass: $m > 2-3 \times 10^{-21}$~eV~\cite{Armengaud:2017nkf, Irsic:2017yje}. The latter bound is applicable to our case as well. 

One important remark is in order here. Having a relatively large coupling constant $\beta$ appears to be advantageous from the viewpoint of testability of the model. 
On the other hand, this may lead to thermalization of DM through the inflaton coupled to the Standard Model fields produced in the course of (p)reheating. Thermalized DM with the 
mass $M \simeq 10^{-21}$~eV behaves as hot DM and can
constitute at best a few percent of the total DM in the
Universe. Another issue is related to the
stability of the produced component (DM and baryons) with
respect to decay into light Standard Model particles through the higher order
operators induced by the coupling to inflaton. This study is out of the scope of the present work.

\section{Discussion}

In this work we suggested a mechanism to generate DM/BA from inflation.
We focused on a particular model of a super-heavy complex scalar field with global $U(1)$-symmetry explicitly broken by the direct coupling of its phase to the inflaton. 

It is worth to note that were we only interested in DM generation during inflation, a simpler model with a real scalar field 
$\Phi$ would suffice.
Indeed, once a linear coupling is introduced, $\sim \Phi\cdot T_{infl}$ (cf. Eq.~\eqref{interaction}), the source term appears in the equation for the scalar field. 
The latter produces the non-zero expectation value of the field $\Phi$ during inflation. Upon the inflaton decay, the vacuum value again takes the zeroth value. 
The scalar $\Phi$ then oscillates about its minimum with the amplitude set by its expectation value during inflation. 

Nevertheless, the model of the complex scalar field is more advantageous because it provides a link to BA as well. 
The latter can be generated using the same mechanism, which leads to the DM production. In this case, the complex field $\Psi$ must carry non-zero baryon charge. 
After inflation, the field $\Psi$ eventually decays converting its Noether charge to the Standard Model quark sector. The resulting BA of the Universe is given by $\Delta_{B} \sim \frac{Q}{s}$. 
The mechanisms of the Noether charge transfer to BA are rather generic and not specific for our particular model. 

In the present work, we did not discuss possible signatures of the mechanism proposed in the current and future data. 
The observational signatures depend on particular couplings 
of the complex scalar to Standard Model fields. Typically, the
couplings to the inflaton make DM particles and baryons unstable, though
with the lifetimes largely exceeding the age of the Universe. Hence, if we insist on the super-heavy DM, these
couplings may lead to potentially observable flux of Ultra High Energy Cosmic Rays. 
See Refs.~\cite{Marzola:2016hyt} and~\cite{Kalashev:2017ijd} for the state of art. 
On the other hand, the constraint on the mass $M$ can be significantly relaxed, upon introducing the non-minimal 
coupling to gravity (see the discussion in Section~5). If the mass $M$
is in the ultra-light range, the evolution of halos may be considerably affected compared to the Cold DM predictions. 
This is a universal prediction for all the ultra-light DM models. Whether our mechanism of DM
production implies some specific signatures (e.g., due to the induced self-coupling from the exchange by the virtual inflaton), which can tested with observations of the cosmic structures at small scales, remains to be clarified.

\section*{Acknowledgments}
E.B. acknowledges support from PRC CNRS/RFBR (2018--2020) n\textsuperscript{o}1985 ``Gravit\'e modifi\'ee et trous noirs: signatures exp\'erimentales et mod\`eles consistants'',
and from the research program ``Programme national de cosmologie et galaxies'' of the CNRS/INSU, France.
The investigation of the baryon asymmetry generation (D.G.) is supported by the RSF grant 14-22-00161. 
The work of S.R. is supported by the funds from the European Regional Development Fund and the Czech Ministry of Education, Youth and Sports (M\v SMT): Project CoGraDS-CZ.02.1.01/0.0/0.0/15\_003/0000437. 

\section*{Appendix: Light DM ($M \ll H$) minimally coupled to gravity}

\subsection*{Non-interacting case}

In this Appendix, we discuss DM composed of the complex scalar field $\Psi$ with the mass $M$ much smaller than the Hubble rate $H$ at inflation, $M \ll H$. We start with the non-interacting case. 
In this situation, the Noether charge density $Q=\lambda^2 \dot{\varphi}$ equals zero, $Q=0$. Hence, $\dot{\varphi}=0$, and the 
complex field evolves only in the radial direction. Its evolution is thus indistinguishable from that of the real scalar field. Namely, it slowly rolls 
during inflation. At the hot epoch, as the Hubble rate drops substantially, so that $H (t_{osc}) \sim M$, the standard dust-like evolution begins; 
$t_{osc}$ is the time, when the field $\Psi$ starts oscillating.

During inflation, the amplitude $\lambda$ acquires isocurvature perturbations, which remain frozen beyond the horizon at the value $\delta \lambda_{iso} \simeq \frac{H}{2\pi}$ until the onset of oscillations. 
At the hot radiation-dominated stage, the DM isocurvature perturbation is defined as 
\begin{equation}
\nonumber 
{\cal S}_{DM}\equiv 3H \left( \frac{\delta \rho_{DM}}{\dot{\rho}_{DM}}-\frac{\delta \rho_{\gamma}}{\dot{\rho}_{\gamma}}\right)=3H\frac{\delta \rho_{DM, iso}}{\dot{\rho}_{DM}} \; .
\end{equation}
Here $\rho_{DM}$ and $\rho_{\gamma}$ are the DM and radiation energy density; 
$\delta \rho_{DM}$ and $\delta \rho_{\gamma}$ are their perturbations, respectively. In the second equality above we use the fact that the isocurvature perturbation of ultra-relativistic 
matter equals zero during radiation-dominated stage (by definition). Using that $\dot{\rho}_{DM}+3H \rho_{DM}=0$ during oscillations, one gets for the power spectrum of DM isocurvature perturbations, 
\begin{equation}
\nonumber
{\cal P}_{{\cal S_{DM}}} =  \langle \left( \frac{\delta \rho_{DM, iso}}{\rho_{DM}} \right)^2\rangle   \; . 
\end{equation}
We then substitute $\rho_{DM}=M^2 \lambda^2$ and $\delta \rho_{DM} =2M^2 \lambda \delta \lambda$ (these are the values obtained upon averaging over oscillations). The background value $\lambda$ and the 
perturbation $\delta \lambda_{iso}$ both redshift with time as $\lambda,~\delta \lambda_{iso} \propto \frac{1}{a^{3/2}}$. Important is that their ratio remains constant. Hence, the power spectrum is also constant, and is defined 
by the value of $\lambda$ and the Hubble parameter during inflationary stage,  
\begin{equation}
\label{iso} 
{\cal P}_{{\cal S_{DM}}}=\frac{H^2}{\pi^2 \lambda^2} \; .
\end{equation}

The allowed amount of isocurvature perturbations uncorrelated with adiabatic fluctuations is limited by the Planck data as $\frac{{\cal P}_{\cal S_{DM}}}{{\cal P}_{\cal R}} <0.038$~\cite{Ade:2015lrj}, where
 ${\cal P}_{{\cal R}} = \frac{H^2}{\pi \epsilon M^2_{Pl}}$ is the spectrum of adiabatic perturbations and $\epsilon$ is the slow roll parameter. Thus the amplitude $\lambda$ is bounded as 
\begin{equation}
\label{constrisolambda}
\lambda \gtrsim 10 M_{Pl} \sqrt{ \epsilon} \; . 
\end{equation}
On the other hand, the requirement that the right amount of DM is produced, fixes the amplitude $\lambda$ at the onset of oscillations, 
\begin{equation}
\nonumber 
M^2 \lambda^2 \cdot \frac{T^3_{eq}}{T^3_{osc}} \sim \rho_{rad} (t_{eq})  \; ,
\end{equation}
where the radiation energy density is given by, 
\begin{equation}
\nonumber 
\rho_{rad} (t_{eq})=\frac{g_* (T_{eq}) \pi^2 T^4_{eq}}{30} \; .
\end{equation}
Here $'eq'$ denotes the matter-radiation equality, $T_{eq}$ is the corresponding temperature, $T_{eq} \simeq 1~\mbox{eV}$, and $g_*$ is the number of ultra-relativistic degrees 
of freedom, $g_{*} (T_{eq}) \simeq 10$; $T_{osc}$ is the temperature of the Universe at the time, when the scalar field starts to oscillate. This occurs, when the Hubble rate is of the order of the mass $M$, i.e., $H (T_{osc}) \sim M$. 
Hence, $T_{osc} \simeq \frac{\sqrt{M M_{Pl}}}{g^{1/4}_*}$, where we make use of the relation between the Hubble rate and temperature at the radiation dominated stage, $H(T) \approx 1.66 \sqrt{g_* (T)} \frac{T^2}{M_{Pl}}$. We conclude with the following estimate for the field $\Psi$ amplitude at the onset of oscillations \footnote{We ignore the possible difference in the number of ultra-relativistic degrees of freedom at the temperatures $T_{osc}$ and $T_{eq}$.}: 
\begin{equation}
\label{estlambdaonset}
\lambda  \simeq \frac{\pi}{\sqrt{30}} \frac{\sqrt{T_{eq} \cdot M_{Pl}}}{{g}^{3/8}_*} \left(\frac{M_{Pl}}{M} \right)^{1/4} \; .
\end{equation} 
Together with the constraint~\eqref{constrisolambda}, this severely limits the range of allowed masses $M$, i.e., $M \lesssim \frac{10^{-34}\text{eV}}{3\epsilon^2}$. In particular, for $M \gtrsim 10^{-21}$~eV, one gets $\epsilon \lesssim 2 \cdot 10^{-7}$. 
While these are the crude estimates, the message is clear: DM production 
is possible only for very light masses $M$ and at inflation with low
energy scale. 

\subsection*{General case}

 \begin{figure}[tb!]
\begin{center}
\includegraphics[width=0.8\columnwidth,angle=0]{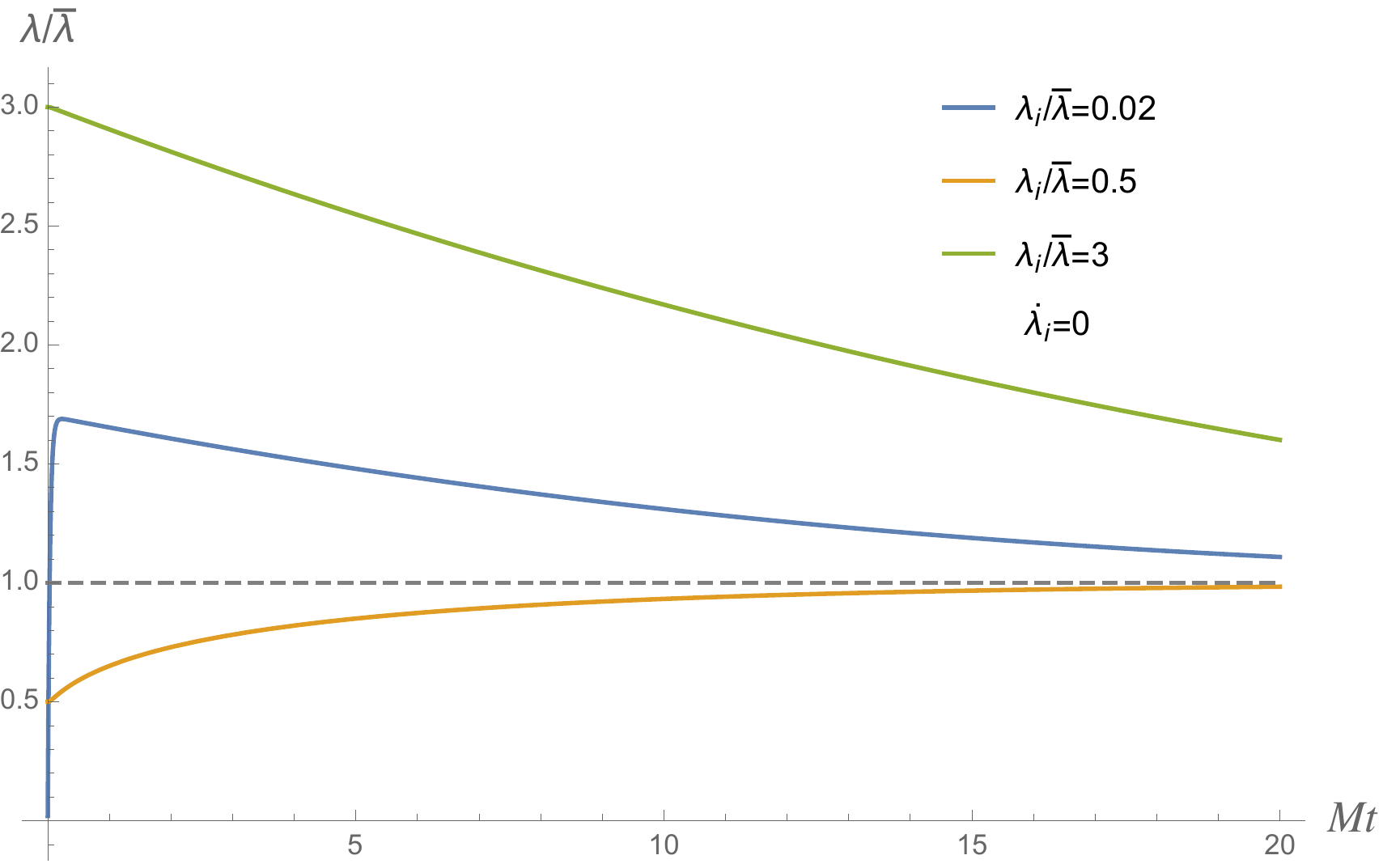}
\caption{Evolution of the field $\lambda$ is shown in the small mass regime $M \ll H$. We have chosen $\frac{M}{H}=0.1$. The approximation of the de Sitter space-time is used. Independently of the initial value $\lambda_i$, the amplitude $\lambda$ 
is always in the slow roll regime during inflation (green and orange lines) or quickly ends up slowly rolling its effective potential (blue line).}\label{smallmass}
\end{center}
\end{figure}
Now let us switch to the case, when the light complex scalar interacts with the inflaton through the phase. 
Again, as in the non-interacting case considered above, the amplitude $\lambda$ slowly rolls (or ends up slowly rolling) the effective potential~\eqref{effectivepotential}, see Fig.~\ref{smallmass}. 
This fact alone sets the lower bound on the physically relevant values of the amplitude $\lambda$.  Let us show this explicitly. In the slow roll regime, the equation for the amplitude simplifies to 
\begin{equation}
\label{sr}
3H \dot{\lambda}=\frac{Q^2}{\lambda^3} -M^2 \lambda\; ,
\end{equation}
where the Noether charge density is still given by Eq.~\eqref{noethercharge}. If $\lambda \gg \bar{\lambda}$, the first term on the r.h.s. is negligible, 
and the dynamics of the field $\lambda$ is determined by the mass term. In that case, the slow roll of the amplitude is guaranteed automatically, since we assume $M \ll H$. 
Therefore, we will focus on small field values $\lambda \ll \bar{\lambda}$ and neglect the second term on the r.h.s. of Eq.~\eqref{sr}. In the exact de Sitter space-time approximation the solution of Eq.~\eqref{sr} reads, 
\begin{equation}
\label{slowr}
\lambda=\left(\frac{4Q^2}{3H}t+C \right)^{1/4} \; ,
\end{equation}
where $C$ is the integration constant. The slow roll approximation is valid provided that the following condition is fullfiled,
\begin{equation}
\nonumber 
\left|\frac{\ddot{\lambda}}{3H \dot{\lambda}} \right|\ll 1 \; .
\end{equation}
Once this condition is satisfied at some time $t$, it becomes progressively better later on. 
Substituting the solution~\eqref{slowr}, we can rewrite the latter inequality as 
\begin{equation}
\nonumber 
\frac{Q^2}{H^2}\ll \frac{4Q^2}{3H}t+C \; .
\end{equation}
If $\frac{Q^2}{H}t \gg C$, one gets $Ht \gg 1$. Substituting this condition into the solution~\eqref{slowr}, we obtain the desired bound
\begin{equation}
\label{ineq}
\sqrt{\frac{Q}{H}} \ll \lambda  \; . 
\end{equation}
One can check that the same inequality follows, if $C \gg \frac{Q^2}{H}t$. This lower bound translates into the upper bound on the phase velocity,
\begin{equation}
\label{upperbound}
 \dot{\varphi} \ll H \; .
\end{equation} 

Following the inflaton decay, the complex scalar (now, a free field) continues the slow roll until the point during the radiation-dominated stage, when $H \sim M$. 
Since then, it experiences oscillations and cosmologically behaves as dust. The constant $\beta$ can be constrained from the observed relic abundance of DM today and the inequality~\eqref{ineq}. The result reads, 
\begin{equation}
\nonumber 
\beta \ll \frac{T_{eq}}{\sqrt{M \cdot M_{Pl}}} \; .
\end{equation}
Note that the scenario with light DM allows for much larger values of the constant $\beta$. So, for $M \sim 1$ eV, it can be as large as $\beta \sim 10^{-14}$. 
From the viewpoint of testability of the model, this may be considered as an advantage over the case 
of the super-heavy DM, where the coupling constant as small as $\beta \sim 10^{-26}$ is required. 

Despite this advantage, the case of light DM is plagued by large isocurvature perturbations. Indeed, the latter for the field $\lambda$ 
are determined by the effective mass~\eqref{effectivemass}. Given the upper bound~\eqref{upperbound} and our assumption $M \ll H$, one gets $M_{eff} \ll H$. 
Consequently, isocurvature perturbations of the field $\lambda$ are those of the massless scalar field, and can be estimated by $\delta \lambda_{iso} \simeq \frac{H}{2\pi}$. 
It is then straightforward to show that the spectrum of isocurvature perturbations can be estimated by Eq.~\eqref{iso}. The rest of estimates 
is also the same as in the non-interacting case considered in the first part of the Appendix. To conclude, the case of light DM is excluded by the current cosmological observations, unless 
the scale of inflation and/or the mass $M$ are very low.

\end{document}